\documentstyle[12pt,epsf]{article}
\newcommand{\beq}{\begin{equation}}
\newcommand{\eeq}{\end{equation}}
\begin{document}
\begin{center}

{\bf Quasiparticles in a strongly correlated liquid with the fermion
condensate: applications to high-temperature superconductors}

\vspace{0.2cm}

S.A. Artamonov and V.R. Shaginyan\footnote{E--mail:
vrshag@thd.pnpi.spb.ru}\\

Petersburg Nuclear Physics Institute, Russian Academy of
Sciences, Gatchina, 188350 Russia\\

\end{center}
\begin{abstract}
A model of a strongly correlated electron liquid based on the
fermion condensation (FC) is extended to high-temperature
superconductors. Within our model, the appearance of FC presents a
boundary separating the region of a strongly interacting electron
liquid from the region of a strongly correlated electron liquid. We
study the superconductivity of a strongly correlated liquid and show
that under certain conditions, the superconductivity vanishes at
temperatures $T>T_c\simeq T_{node}$, with the superconducting gap
being smoothly transformed into a pseudogap. As the result, the
pseudogap occupies only a part of the Fermi surface. The gapped area
shrinks with increasing the temperature and vanishes at $T=T^*$. The
single-particle excitation width is also studied.  The quasiparticle
dispersion in systems with FC can be represented by two straight
lines characterized by the respective effective masses $M^*_{FC}$ and
$M^*_L$, and intersecting near the binding energy that
is of the order of the superconducting gap. It is argued that this
strong change of the quasiparticle dispersion at the binding can be
enhanced in underdoped samples because of strengthening the FC
influence. The FC phase transition in the presence of the
superconductivity is examined, and it is shown that this phase
transition can be considered as kinetic energy driven.
\end{abstract}


\section{Introduction}
Unusual properties of the normal state of high-temperature
superconductors have been attracting attention for a long time.
In describing these properties, which are well beyond the
standard Fermi liquid theory, the notion of a strongly correlated
liquid has emerged (see e.g. \cite{sd,ift}). Later on,
angle-resolved photoemission studies revealed unusual properties
observed in underdoped samples, with the leading edge gap
discovered up to the temperature $T^*>T_c$. This behavior is
interpreted as coming from the pseudogap formation; it was observed in
a number of underdoped compounds such as
YBa$_2$Cu$_3$O$_{6+x}$, Bi$_2$Sr$_2$CaCu$_2$O$_{8+\delta}$,
etc. As $T$ increases above $T^*$, the pseudogap closes, leading to
a large Fermi surface and an extremely flat dispersion in electronic
spectra, which is called the extended Van Hove singularity $[3-7]$.
Thereupon recent discovery of a new energy scale for quasiparticle
dispersion in Bi$_2$Sr$_2$CaCu$_2$O$_{8+\delta}$ in the
superconducting and in normal states \cite{blk,krc} have brought new
insight to the physics of high temperature superconductors (HTS),
imposing serious constraints upon the theory of HTS.
The newly discovered additional energy scale manifests itself as a
break in the quasiparticle dispersion near $(50-70)\,$meV, which
results in a drastic change of the quasiparticle velocity
\cite{blk,krc,vall}. Such a behavior is definitely different from
what one could expect in a normal Fermi liquid.

A correlated liquid can be described in conventional terms,
assuming that the correlated regime is related  with the
noninteracting Fermi gas by adiabatic continuity. This is done in
the well-known Landau theory of the normal Fermi liquid,
but the question arising at this point is whether this is possible.
Most likely, the answer is negative. To tackle the
above-mentioned problems, we consider a model
where a strongly correlated liquid is separated from the
conventional Fermi liquid by a phase transition related to the onset
of the FC \cite{ks,vol}. The purpose of our paper is to show that
without any adjustable parameters, a number of fundamental
problems of strongly correlated systems are naturally explained
within the model. The paper is organized as follows. In Sec. 2, we
consider the general features of Fermi systems with the FC.
In Sec. 3, we show that the pseudogap behavior can be
understood within the standard BCS superconductivity mechanism
provided the appearance of FC is taken into account. In Sec. 4,
we analyze the condensation energy that is liberated when
the system in question undergoes the superconducting phase transition
superimposing on the FC phase transition. In Sec. 5, we describe
the quasiparticle dispersion and lineshape. Finally, in Sec. 6, we
summarize our main results.

\section{The main features of liquids with FC}
We first consider the key points of the FC theory.
The FC is related to a new class of solutions of
the Fermi liquid theory equation \cite{lan}
\beq \frac{\delta (F-\mu N)}{\delta
n(p,T)}=\varepsilon(p,T)-\mu(T)-T\ln\frac{1-n(p,T)}{n(p,T)}=0\eeq
for the quasiparticle distribution function $n(p,T)$ depending on
the momentum $p$ and the temperature $T$. Here $F$ is the free
energy, $\mu$ is the chemical potential, and
$\varepsilon(p,T)=\delta E/\delta n(p,T)$ is the quasiparticle
energy, which is a functional of $n(p,T)$ just like the energy $E$
and the other thermodynamic functions. Equation (1) is usually
represented as the Fermi-Dirac distribution
\beq
n(p,T)=\left\{1+\exp\left[\frac{(\varepsilon(p,T)-\mu)}
{T}\right]\right\}^{-1}.\eeq
In a homogeneous matter and at $T = 0$,
one obtains from Eq. (2) the standard solution
$n_F(p,T=0)=\theta(p_F-p)$, with $\varepsilon(p\simeq
p_F)-\mu=p_F(p-p_F)/M^*_L$, where $p_F$ is the Fermi momentum and
$M^*_L$ is the commonly used effective mass \cite{lan},
\beq \frac{1}{M^*_L}
=\frac{1}{p}\frac{d\varepsilon(p,T=0)}{dp}|_{p=p_F}.\eeq
It is assumed to be positive and finite at the Fermi momentum
$p_F$. This implies the $T$-dependent corrections to $M^*_L$, the
quasiparticle energy $\varepsilon (p)$, and the other quantities
start with $T^2$-terms.

But this solution of Eq. (1) is not the only one possible. There
exist ``anomalous" solutions of Eq. (1) associated
with the so-called fermion condensation \cite{ks,ksk,dkss}. Being
continuous and satisfying the inequality $0<n(p)<1$ within some
region in $p$, such a solution $n(p)$ admits a finite limit for the
logarithm in Eq. (1) as $T\rightarrow 0$, yielding \beq
\varepsilon(p)=\frac{\delta E[n(p)]}{\delta n(p)} =\mu, \: p_i\leq p
\leq p_f. \eeq Equation (4) is used in searching the minimum
value of $E$ as a functional of $n(p)$ under the assumption that a
strong rearrangement of the single-particle spectrum can occur.
We see from Eq. (4) that the occupation numbers $n(p)$ become
variational parameters: the solution $n(p)$ exists if the energy $E$
is decreased by alteration of the occupation numbers. Thus, within the
region $p_i<p<p_f$, the solution $n(p)$ deviates from the Fermi step
function $n_F(p)$ such that the energy $\varepsilon(p)$
stays constant, while $n(p)$ coincides with $n_F(p)$ outside this
region. As a result, the standard Kohn --- Sham scheme for the
single-particle equations is no longer valid beyond
the FC phase transition point \cite{vsl}. This behavior of systems
with the FC is clearly different from what one expects from the well
known local density calculations; therefore, these calculations are
not applicable to systems with the FC. On the other hand, the
quasiparticle formalism is applicable to this problem, because as we
see in what follows, the damping of single-particle excitations is
not large compared to their energy \cite{dkss}. It is also seen from
Eq. (4) that a system with the FC has a well-defined Fermi surface.

It follows from Eq. (1) that
at low $T$, new solutions within the interval occupied by the
fermion condensate have the spectrum $\varepsilon(p,T)$ that is
linear in T \cite{dkss,kcs},  \beq
\varepsilon(p,T)-\mu(T)\simeq\frac{(p-p_F)p_F}{M^*_{FC}}
\simeq T(1-2n(p))\ll T_f. \eeq
Here $T_f$ is the quasi-FC phase transition
temperature above which FC effects become insignificant \cite{dkss},
\beq
\frac{T_f}{\varepsilon_F}\sim\frac{p_f^2-p_i^2}{2M\varepsilon_F}
\sim\frac{\Omega_{FC}}{\Omega_F},\eeq
where $M$ is the bare electron mass,
$\Omega_{FC}$ is the condensate volume, $\varepsilon_F$
is the Fermi energy, and $\Omega_F$ is the volume of the Fermi
sphere. One can imagine that the dispersionless plateau
$\varepsilon(p)=\mu$ given by Eq. (4) is slightly tilted
counter-clockwise about $\mu$ and rounded off at the end points.  If
$T\ll T_f$, it follows from Eqs. (1) and (5) that the effective mass
$M^*_{FC}$ related to the FC is temperature dependent
\beq \frac{M^*_{FC}}{M} \sim
\frac{N(0)}{N_0(0)}\sim\frac{T_f}{T},\eeq
where $N_0(0)$ is the density of states of
the noninteracting electron gas, and $N(0)$ is the density of
states at the Fermi level. We note that outside the
FC region, the single-particle spectrum is not distinctly affected by
temperature, being determined by the effective mass $M^*_L$ given by
Eq. (3), which is now evaluated at $p\leq p_i$. Thus, we are led to
the conclusion that systems with a FC must be characterized by two
effective masses:  $M^*_{FC}$ related to the single particle spectrum
of a low-energy scale and $M^*_L$ related to the spectrum of a higher
energy scale. The existence of these two effective masses can be
observed as a break in the quasiparticle dispersion. This break is
observed at temperatures $T\ll T_f$, and also when the
superconducting state is superimposed on the FC state. In
the former case, the occupation numbers over the area occupied by the
fermion condensate are slightly disturbed by the pairing correlations
such that the effective mass $M^*_{FC}$ becomes large but finite. We
remark that at comparatively low temperatures, the FC and
superconductivity go together because of the remarkable peculiarities
of the FC phase transition. This transition is related to
a spontaneous gauge symmetry breaking: the superconductivity
order parameter $\kappa(p)=\sqrt{n(p)[1-n(p)]}$ has a nonzero value
over the region occupied by the fermion condensate, while the gap
$\Delta$ can vanish \cite{dkss,vsl}.

It is seen from Eq. (4) that at the FC phase transition point
$p_f\to p_i\to p_F$, while the effective mass and the density of
states  tend to the infinity as follows from Eqs. (4) and (7). One
can conclude that the beginning of the FC phase transition is
related to the absolute growth of $M^*_{FC}$. The onset of the
charge density-wave instability in an electron system,
which occurs as soon as the effective electron-electron
interaction constant $r_s$ reaches its critical value $r_{cdw}$,
must be preceded by the unbounded growth of the effective mass
\cite{ksz}. For a simple electron liquid, the effective constant is
proportional to the dimensionless average distance
$r_s\sim r_0/a_B$ between particles of the system in question,
with $r_0$ being the average distance and $a_B$ the Bohr radius.
The physical reason for this growth is the contribution of the
virtual charge density fluctuations to the effective mass. The
excitation energy of these fluctuations becomes very small if
$r_s\simeq r_{cdw}$. Thus, a FC can occur when $r_s\sim r_{cdw}$.
The standard Fermi liquid behavior can therefore be
broken by strong charge fluctuations when the insulator regime
is approached in a continuous fashion. We recall that the
charge-density wave instability occurs in three-dimensional
\cite{lp} and two-dimensional (2D) electron liquids \cite{sns} at a
sufficiently high $r_s$. As soon as $r_s$ reaches its critical value
$r_{FC}<r_{cdw}$, the FC phase transition occurs. Thereafter, the
condensate volume is proportional to $r_s-r_{FC}$ and also
$T_f/\varepsilon_F\sim r_s-r_{FC}$ \cite{dkss,ksz}. In fact, the
effective coupling constant $r_s$ increases with decreasing doping.
It is assumed that both $T_f$ and condensate volume
$\Omega_{FC}$ build up with decreasing doping. The FC then serves as a
stimulating source of new phase transitions lifting the degeneracy
of the spectrum. The FC can produce, for instance, the spin density
wave (SDW) phase transition or the antiferromagnetic one, thereby
promoting a variety of the system properties. We note
that the SDW phase transition, the antiferromagnetic transition,
and the charge density one also depend on $r_s$ and occur
at a sufficiently large value of $r_s$ even if the FC is absent.
The superconducting phase transition is also aided by the FC.  We
analyze the situation where the superconductivity wins the
competition with the other phase transitions up to a temperature
$T_c$. Above the temperature $T^*\ll T_f$ the system under
consideration is in its anomalous normal state, Eq. (7) is valid,
and one can observe smooth non-dispersive segments of the spectra
at the Fermi surface \cite{nr}.

\section{Superconductivity in the presence of FC}
We focus our attention on investigating the pseudogap
that is formed above $T_c$ in underdoped (UD) high-temperature
superconductors $[4-8]$. As we see in what follows, the existence of
the pseudogap is closely allied with the presence of the FC
characterized by a sufficiently high temperature $T_f$ given by Eq.
(6). Thus, the pseudogap is peculiar to UD samples, while optimally
doped (OP) and overdoped (OD) samples may not exhibit this feature.
We consider a 2D liquid on a simple square lattice that has a
superconducting state with the d-wave symmetry of the order parameter
$\kappa$. We assume that the long-range component $V_{lr}({\bf q})$
of the particle-particle interaction $V_{pp}({\bf q})$ is
repulsive and has the radius $q_{lr}$ in the momentum space such that
$p_F/q_{lr}\leq 1$.  The short-range component $V_{sr}({\bf q})$ is
relatively large and attractive, with its radius $p_F/q_{sr}\gg 1$.
In agreement with the d-symmetry requirements the low
temperature gap $\Delta$ is then given by the expression
\cite{scal,scalr,abr}
$$\Delta(\phi)=2\kappa(\phi)E(\phi)\simeq \Delta_1
\cos2\phi=\Delta_1(x^2-y^2),$$ where
$E(\phi)=\sqrt{\varepsilon^2(\phi)+\Delta^2(\phi)}$ and
$\Delta_1$ is the maximal gap. At finite temperatures, the equation
for the gap can be written as
\beq\Delta(p,\phi)=-\int_0^{2\pi}\int
V_{pp}(p,\phi,p_1,\phi_1)\kappa(p_1,\phi_1)
\tanh{E(p_1,\phi_1)\over 2T}\frac{p_1dp_1d\phi_1}{4\pi^2},\eeq
where $p$ is the absolute value of the momentum and $\phi$ is
the angle. It is also assumed that the FC arises near the van Hove
singularities, leading to a large density
of states at these points in accordance with Eq. (7). We note that
the different FC areas overlap {\it only} slightly \cite{kcs}.
$\Delta(\phi)$ obeys the following equation that is determined by
the chosen interaction $V_{pp}$
\beq\Delta({\pi\over 4}+\phi)=-\Delta({\pi\over 4}-\phi).\eeq
It vanishes at $\pi/4$ and can therefore be expanded in the
Taylor series around $\pi/4$, with $p\approx p_F$:
\beq\Delta(p,\theta)= \theta a - \theta^3 b +...,\eeq where
$\theta = \phi - {\pi \over 4}$.
Hereafter, we consider solutions of
Eq. (8) on the interval $0<\theta<\pi/4$. We transform Eq. (8) by
setting $p\approx p_F$ and separating the contribution $I_{lr}$
coming from $V_{lr}$, with the contribution related to $V_{sr}$
denoted by $I_{sr}$. At small angles, $I_{lr}$ can be approximated
in accordance with (10) by $I_{lr}=\theta A +\theta^3 B$, with
the parameters $A$ and $B$ independent of $T$ if $T\leq T^*\ll T_f$,
because they are defined by the integral over the regions occupied
by the FC. This theoretical observation is consistent with the
experimental results showing that $\Delta_1$ is essentially
$T$-independent at the temperatures $T<T^*$ \cite{nr}. The
coefficients of the expansion of $I_{sr}$ in powers of
$\theta$ depend on $T$. It is therefore more
convenient to use the integral representation for $I_{sr}$
following from (8). We, thus, have
\beq\Delta(\theta)=I_{sr} +I_{lr}=-\int_0^{2\pi}\int
V_{sr}(\theta,p_1,\phi_1)\kappa(p_1,\phi_1)
\tanh{E(p_1,\phi_1)\over 2T}\frac{p_1dp_1d\phi_1}{4\pi^2}+\theta
A+\theta^3 B.\eeq In Eq. (11), the variable $p$ was omitted since
$p \approx p_F$. It is seen from this equation that the
 FC produces the free term $\theta A+\theta^3 B$.
In what follows, we show that at $T\geq T_{node}$, the solution of
Eq. (11) has the second node at $\theta_c(T)$ in the vicinity of the
first node at $\pi/4$. We also demonstrate that the temperature
$T_{node}$ has the meaning of the temperature $T_c$ at which the
superconductivity vanishes.  To show this, we simplify Eq. (11)
to an algebraic equation. We have $I_{sr}\sim (V_0/T)\theta$
because $\tanh(E/2T)\approx E/2T$ for $E\ll T$ and $T\approx
T_{node}$, as is the case in the vicinity of the gap node at
$\theta=0$. The integration in Eq. (11) runs over a small
area located at the gap node because of the small radius of $V_{sr}$.
Dividing both parts of Eq. (11) by $\kappa(\theta)$, we obtain
\beq E(\theta) =-({V_0\over T}-A_1-\theta^2 B_1)|\theta|, \eeq
where $A_1$ and $B_1$ are new
constants and $V_0\sim V_{sr}(0)$ is a constant.
Imposing the condition that Eq. (8)
has the only solution $\Delta\equiv 0$ when $V_{sr}=0$,
we see that $A_1$ is negative and  $B_1$
is positive. The factor in the brackets on the right-hand side of
Eq.  (12) changes its sign at some temperature $T_{node}\approx
V_0/A_1$; on the other hand, the excitation energy must be
$E(\theta)>0$. Therefore, we have two possibilities \cite{avl,vs}.
The first follows from the assumption that $\Delta(\theta)\equiv 0$
if $\theta$ belongs to the interval $\Omega_n$ $[0<\theta<\theta_c]$.
In this case, for $T>T_{node} $ we must solve Eq. (8) with the
condition
$$\Delta(\theta)\equiv 0,\,\,\, {\rm if:}\,\,
0<\theta<\theta_c;\,\,\, T_{node}<T.$$ This resembles Eq. (4) with
the parameter $\mu$ being equal to zero. The similarity is not
coincidental, because we are searching for new
solutions in both cases.  Such solutions do exist because the points
$\theta=0$ and $\theta=\theta_c$ represent the branching points of
the solutions.
The second possibility can occur if the
above solution does not lead to a minimum value of the
free energy. Because the excitation energy must
be positive for a stable state, the sign
of $\Delta$ must be reversed at the point $\theta=\theta_c$. Then
the gap $\Delta(\theta)$ has the same sign within the  interval
$\Omega_n$ and changes its sign once more at the point $\theta=0$,
with $\Delta(\theta_c)=\Delta(0)=0$.  Thus, we conclude that the gap
$\Delta$ possesses new nodes at $T>T_{node}$ \cite{vs}, see Fig.
1. It can be seen from Eq. (12) that the angle $\theta_c$ is related
to $T>T_{node}$ by  \beq
T\approx\frac{V_0}{(A_1+B_1\theta_c^2)}.\eeq It follows from the
above consideration and Eq. (12) that even below $T_{node}$, the
order parameter $\Delta$  cannot be approximated by a simple d-wave
form; a more sophisticated expression must be used to fit
the flattening of the gap $\Delta$ around the node.
The following expression can be used for this purpose,
\beq \Delta(\phi)=\Delta_1[B\cos(2\phi)+(1-B)\cos(6\phi)]. \eeq
Here $0<B<1$ in accordance with the experimental results \cite{mn}
and the term involving $\cos (6\phi)$ is the next
compatible with the d-symmetry of the gap. It also follows from Eq.
(12) that the parameter $B$ is a decreasing function of the
temperature. At the temperatures $T>T_{node}$, the value of $(1-B)$
is sufficiently large to produce new nodes of $\Delta$ given by Eq.
(14).

As an example of the solutions of Eqs. (8) and (11), we show,
in Fig. 1 the gap $\Delta(\phi)$ calculated at three different
temperatures $0.9\,T_{node}$, $T_{node}$, and $1.2\,T_{node}$.
An important difference between curves (b) and (a)
is the flattening of curve (b) at the nodes localized within
the region $\Omega_n$ containing the interval
$-\theta_c\leq\theta\leq\theta_c.$ As seen from Fig.  1,
the flattening occurs as the result of the new nodes restricting the
area $\Omega_n$. It is also seen
from Fig. 1 that the gap $\Delta$ is extremely small over the range
$\Omega_n$.

\begin{figure}[t]
\centerline{\epsfxsize=16cm
\epsfbox{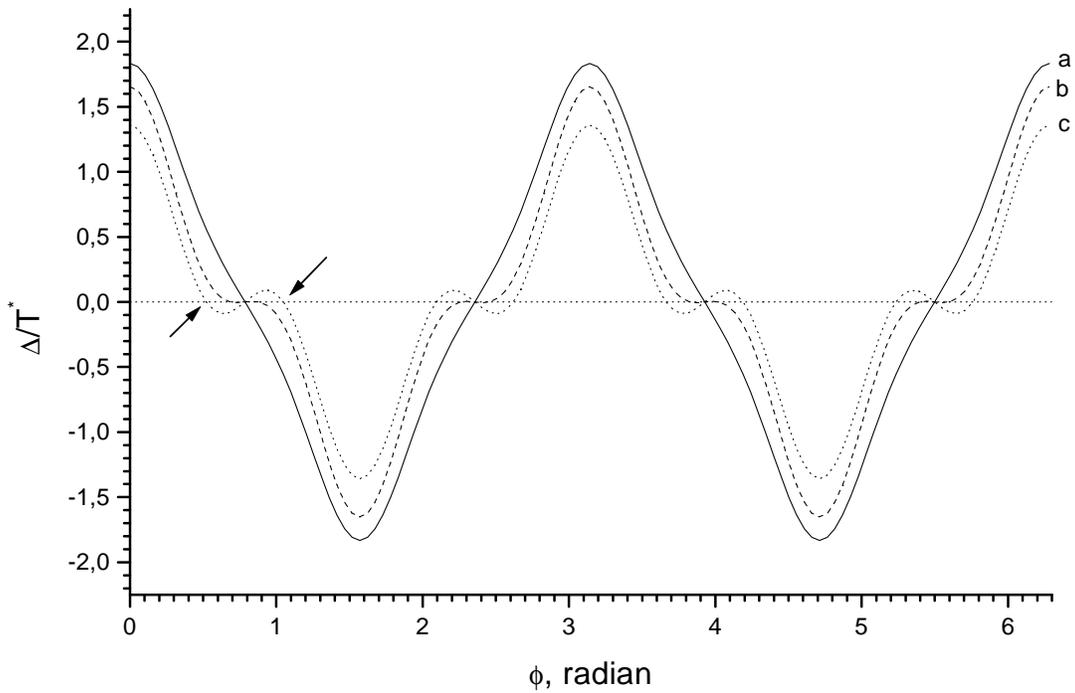}}
\caption{The gap $\Delta$
as a function of $\phi$ calculated at three different temperatures
expressed in terms of $T_{node}\simeq T_c$, while $\Delta$ is
presented in terms of $T^*$. Curve (a), solid line, shows the gap
calculated at temperature $0.9 T_{node}$. In curve (b), dashed line,
the gap is given at $T_{node}$.  Note the important difference in
curve (b) compared with curve (a) due to a flattening of the curve
(b) over the region $\Omega_n$. Calculated $\Delta(\phi)$ at $1.2
T_{node}$ is shown by curve (c), dotted line. The arrows indicate the
two nodes restricting area $\Omega_n$ and emerged at $T_{node}$.}
\end{figure}

It was recently shown in a number of papers (see, e.g.,
\cite{nm,th}) that there exists an interplay between the magnetism
and the superconductivity order parameters, leading to the damping of
the magnetism order parameter below $T_c$.  Conversely, one can
anticipate the damping of the superconductivity order parameter by
magnetism. Thus, we conclude that the gap in the range $\Omega_n$
can be destroyed by strong antiferromagnetic correlations (or by spin
density waves) existing in underdoped superconductors
\cite{sp,pr}. It is believed that impurities can easily destroy
$\Delta$ in the considered area. As a result, one is led to the
conclusion that $T_c\simeq T_{node}$, with the exact value of $T_c$
defined by the competition between the antiferromagnetic
correlations (or spin density waves) and the superconducting
correlations over the range $\Omega_n$.

We now consider the possibility for two quite different
properties, the superconductivity and static spin density wave
(SDW), to coexist. We start by briefly outlining the main
features of the SDW \cite{ov}. A simple example is given by the
linear SDW, with the net spin
polarization ${\bf P}(\bf r)$
\beq {\bf P}({\bf r})=P_0 {\bf e}\cos(\hat{Qx}), \eeq
where $\hat{Qx}$ is the angle between
the vectors ${\bf Q}$ and ${\bf x}$. For convenience, the direction of
the SDW is taken along the ${\bf x}$ axis,
and ${\bf e}$ is the unit polarization vector, which in general
can have any orientation with respect to ${\bf Q}$. In
contrast to the superconductivity, SDW can occupy only a part of the
Fermi sphere with the volume $\delta S\simeq p_F\delta\phi\delta k$,
where $\delta\phi$ is the Fermi surface angle and $\delta k$ is the
``penetration depth" of the SDW into the Fermi sphere.  At $T=0$,
the energy gain $\delta W$ due to the onset of SDW is given by
\beq \delta W\simeq g^2 N(0)\delta\phi, \eeq where $g$
is the SDW gap determined by the formula \cite{ov}
\beq g\simeq \frac{p_F\delta k}{N(0)}
\exp(-\frac{4}{N(0)\gamma_0\delta\phi}),\eeq where $\gamma_0$ is the
coupling constant. As seen from Eq. (8), the variation of
the gap within some area produces a variation of the gap over
the entire occupied area with the same order of magnitude. Therefore,
elimination of $\Delta$ over a segment $\delta\phi$ requires the
energy $\delta E_1\sim N(0)\Delta^2(\phi)$. We conclude that
at $T<T_{node}$, the destruction of the gap on the interval
$\delta\phi$ eliminates $\Delta$ over the entire region, because
$\delta E_1$ is comparable with the gain $\delta E$ due to the
superconducting state. A different situation occurs at the
temperatures $T>T_{node}$, when $\Delta$ is extremely small in
$\Omega_n$ and the corresponding destruction energy satisfies
inequality $\delta E_1\ll\delta E$. Equations (16) and (17) are very
similar to the corresponding BCS equations and this similarity also
remains at finite temperatures \cite{ov}. Thus, the gain $\delta
W$ and the gap $g$ vary with the temperature similarly
to the superconducting gain $\delta E$ and the gap $\Delta$. We
also assume that the SDW transition temperature $T_n$ is
sufficiently high, namely, $T_n\geq T_c$. We then come to the
conclusion that $\delta E_1<\delta W$, and the region
$\Omega_n$ is therefore occupied by the SDW at temperatures $T\geq
T_{node}$, resulting in the destruction of the superconductivity
\cite{avl,vs}. We note that the Fermi surface angle $\delta\phi$
must be sufficiently large, because the
gap $g$ depends exponentially on $\delta\phi$ in accordance with Eq.
(17). On the other hand, because we are dealing with SDW, we have
$\delta\phi/\pi\sim 10^{-2}$ \cite{ov}. We thus conclude that  a
strong variation of the superconductivity characteristics may be
observed in the vicinity of $T_{node}$.

It follows from the above considerations
that $\Delta(\theta)$ can be destroyed
only locally within the region $\Omega_n$ because of the different
reasons.  It also follows that $T_{node}$ is the temperature at
which the superconductivity vanishes, that is, $T_c\simeq T_{node}$.
As to the gap at $T>T_c$, or more precisely, the pseudogap, it
persists outside the $\Omega_n$ region. In accordance with
\cite{d2,mn}, we see that the superconducting gap $\Delta(\theta)$
smoothly transforms into the pseudogap at $T>T_c$.  We
can therefore expect a dramatic reduction in the difference
between the free energy of the normal and the superconducting state
at $T=T_c$ (the so-called condensation energy, which we consider in
some detail in the next section). It can then be concluded that the
temperature $T^*$ has the physical meaning of the BCS transition
temperature between the state with the order parameter $\kappa\neq 0$
and the normal state. Because $T_c\simeq V_0/A_1$, we find from Eq.
(13) that $\theta_c\sim\sqrt{(T-T_c)/T_c}$. This result is in harmony
with our calculations of the function $\theta_c([T-T_c]/T_c)$ plotted
in Fig. 2. Thus, we conclude that the pseudogap ``dies out" in UD
samples as the temperature $T^*$ is approached. Quite naturally, one
has to recognize that $\Delta_1$ must scale with $T^*$.

\begin{figure}[tbp]
\centerline{\epsfxsize=12cm
\epsfbox{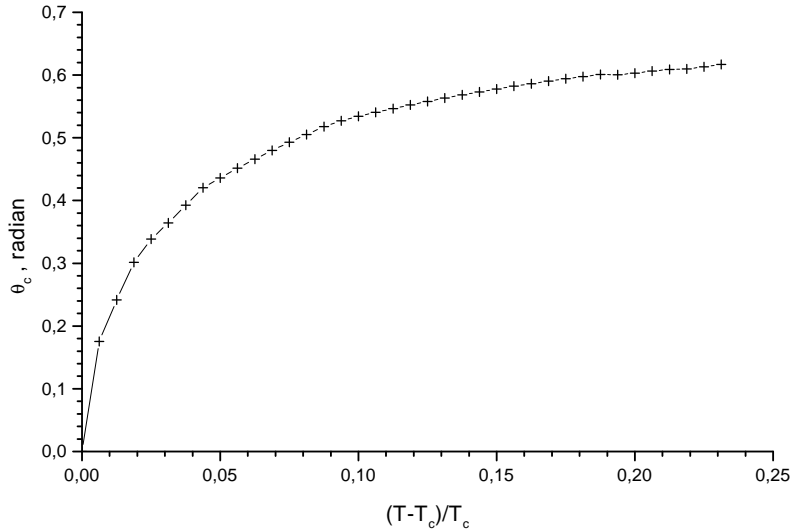}}
\caption{Calculated angle $\theta_c$, pulling apart
the two nodes, as a function of $(T-T_c)/T_c$.}
\end{figure}

A few remarks are in order at this point. On
the basis of the previous consideration, we conclude that the BCS
approach is fruitful in considering
OD, OP, and UD samples in the weak coupling regime.
With more underdoping, the antiferromagnetic correlations become
stronger, breaking down the gap over the range $\Omega_n$ at lower
temperatures. Thus, one observes the decrease of $T_c$ with the
decrease of doping.  On the other hand, the
condensate volume $\Omega_{FC}$ becomes larger with the decrease of
doping, leading to increase of the gap $\Delta_1$ which is
proportional to the volume and interaction $V_{pp}$ \cite{ks}.
Consequently, the temperature $T^*$ becomes higher with
decreasing doping. All these results are in agreement
with the experimental findings \cite{d2,mn}. A peak was
observed at 41 meV$\simeq 2\Delta_1$ in inelastic neutron scattering
from single crystals of the OD, OP, and UD
samples YBa$_2$Cu$_3$O$_{6+x}$ and Bi$_2$Sr$_2$CaCu$_2$O$_{8+\delta}$
at temperatures below $T_c$, while a broad maximum above
$T_c$ exists in underdoped samples only \cite{fong,hhe}. The
explanation of this peak given in \cite{ab} was based on the ideas of
the BCS theory. From the above discussion, it appears that the
same explanation holds for the broad maximum in underdoped samples
above $T_c$ because the physics of the process is essentially the
same.

\section{Condensation energy}
We now consider the energy gain or condensation energy $E_{cond}$
liberated when the system in question undergoes the
superconducting phase transition involved in the FC phase transition.
We set $T=0$ for simplicity. The energy $E_{cond}$ can be
schematically broken into two parts related to the kinetic and
the potential energy. The condensation energy was
considered in \cite{ncon}, where it was argued that the main
contribution to the condensation energy comes from the kinetic
energy, i.e., the superconducting phase transition of
high-temperature superconductors is kinetic energy driven. Here, we
give a possible interpretation of the situation. It is known
\cite{ch} that in the superconducting phase transition,
the positive contribution comes from the potential energy, while the
gain in the kinetic energy is negative. In the other words, the
superconducting phase transition is driven by the gain in the
potential energy. This result is rather obvious because the ground
state energy $E_{gs}$ is given by
\beq E_{gs}[\kappa({\bf p})]=E[n({\bf p})]+
E_{sc}[\kappa({\bf p})], \eeq
with the occupation numbers $n({\bf p})$ determined by
$\kappa({\bf p})=\sqrt{n({\bf p})[1-n({\bf p})]}$.
The second term $E_{sc}[\kappa({\bf p})]$ on the right-hand side of
Eq.  (18) is defined by the superconducting contribution, which in
the simplest case is of the form
\beq E_{sc}[\kappa({\bf p})]=g_2\int V_{pp}({\bf p}_1,{\bf p}_2)
\kappa({\bf p}_1)\kappa({\bf p}_2)
\frac{d{\bf p}_1d{\bf p}_2}{(2\pi)^4}.  \eeq
The first term $E[n({\bf p})]$ can
be taken as \beq E[n({\bf p})]=\int \frac{p^2}{2M} n({\bf
p})\frac{d{\bf p}}{4\pi^2}+ \frac{g_1}{2}\int V({\bf p}_1,{\bf p}_2)
n({\bf p}_1)n({\bf p}_2)
\frac{d{\bf p}_1d{\bf p}_2}{(2\pi)^4}, \eeq
with the second integral playing the
role of the exchange-correlation contribution to the ground state
energy. If the effective mass $M^*_L$ given by Eq. (3) is positive
and finite, $E[n({\bf p})]$ reaches its minimum at $n(p)=n_F(p)$
and increases with the deviation of $n(p)$ from the Fermi
distribution, as it occurs in the presence of superconducting
correlations.  Thus, the standard situation is that the
superconducting phase transition is driven by a decrease of the
potential energy \cite{ch}.  The situation can be different if the
system undergoes the FC phase transition. To see this we
temporarily assume that $g_2\to0$ and rewrite Eq. (20) as
\beq E[n({\bf p})]=\int \varepsilon({\bf p}) n({\bf p})\frac{d{\bf
p}}{4\pi^2}- \frac{g_1}{2}\int V({\bf p}_1,{\bf p}_2) n({\bf
p}_1)n({\bf p}_2) \frac{d{\bf p}_1d{\bf p}_2}{(2\pi)^4}, \eeq with
the single particle energy
\beq \varepsilon({\bf p})=
\frac{\delta E[n({\bf p})]}{\delta n({\bf p})}.\eeq
The energy $E[n({\bf p})]$ can be lowered
by alteration of $n({\bf p})$ if Eq. (4) has solutions. As the
result, we can write the inequality \cite{ks}
\beq
E_{cond}=E_N-E_{FC}\geq\int \left(\varepsilon({\bf p})
-\mu\right)\delta n({\bf p})
\frac{d{\bf p}}{4\pi^2}\geq 0,\eeq
with $E_N$ being the energy of system in its normal state,
$E_{FC}$ the energy with FC, and the integral taken over the region
occupied by FC. The chemical potential $\mu$ preserves the
conservation of the particle number under the variation $\delta
n({\bf p})$. We assume that the kinetic energy is given by the
first term on the right-hand side of Eq. (21).  It then follows from
Eq. (23) that the kinetic energy can be lowered, and this lowering
is driven by the FC phase transition. It is instructive to illustrate
this by a simple example.
We take $V({\bf p}_1,{\bf p}_2) =g_1\delta({\bf p}_1-{\bf p}_2)$,
then $E_{cond}$ given by Eq.
(23) becomes \beq E_{cond}=\int
\left(\varepsilon_0({\bf p})n_F(p)-\varepsilon({\bf p})n(p)\right)
\frac{d{\bf p}}{4\pi^2}
+\frac{g_1}{2}\int \left(n^2(p)-n_F^2(p)\right)
\frac{d{\bf p}}{4\pi^2},\eeq
with $\varepsilon_0({\bf p})$ being the single particle energy of the
normal ground state.
It is easily verified that the second term on the right-hand side of
Eq. (24), which is related to the potential energy  gain, is
negative. This term can be written as
$$\frac{g_1}{2}\int\left(n^2(p)-n_F^2(p)\right)
\frac{d{\bf p}}{4\pi^2}=
\frac{g_1}{2}\int\left(n(p)-n_F(p)\right)
\left(n(p)+n_F(p)\right)\frac{d{\bf p}}{4\pi^2}.$$
Observing that
$$\int\left(n(p)-n_F(p)\right)\frac{d{\bf p}}{4\pi^2}=0$$
because of the particle number conservation and
taking into account that
$$(n(p)+n_F(p))|_{p\leq p_F}>(n(p)+n_F(p))|_{p_F\leq p},$$
we arrive at the conclusion. The
first term is positive because of inequality (23).
Thus, we are led to the conclusion that the FC phase transition can
be considered as driven by the kinetic energy. We now let the
coupling constant $g_2$ be small, then the gap $\Delta$ is
proportional to $g_2$ \cite{ks}.  The optimum values of the
occupation numbers given by Eq. (4) are disturbed, leading to an
increase of the energy $E[n({\bf p})]$.  The positive gain in the
potential energy given by Eq. (19) is driving the
formation of the superconducting ground state. Because the coupling
constant $g_2$ is sufficiently small, the structure of the system
ground state is defined by the FC, and the superconducting state
is a ``shadow" of the FC under these conditions \cite{dkss}. Then,
the main contribution to $E_{cond}$ comes from the FC phase
transition, and the complex transition (FC plus
superconductivity) is kinetic energy driven \cite{ars}. On the other
hand, in the case where FC is weak compared to the
superconductivity (or is absent), we are dealing with a
pure superconducting phase transition, which is obviously potential
energy driven.

\section{Quasiparticle dispersion and lineshape}
We now discuss the origin of two effective masses
$M^*_L$ and $M^*_{FC}$ occurring in the superconducting state
and leading to a nontrivial quasiparticle dispersion and a
change of the quasiparticle velocity. As we see in what follows,
our results are in a reasonably good agreement with the
experimentally deduced data \cite{vall,blk,krc}.
For simplicity we set $T=0$. Varying $E_{gs}$
given by Eq. (18) with respect to $\alpha_{\bf p}$, we find
\beq \frac{E_{gs}[\alpha_{\bf p}]}
{\delta\alpha_{\bf p}} =(\varepsilon({\bf p})-\mu)
\tanh(2\alpha_{\bf p})+\Delta({\bf p})=0,
\eeq
with $n({\bf p})=\cos^2(\alpha_{\bf p})$,
$\kappa({\bf p})=\sin(\alpha_{\bf p})\cos(\alpha_{\bf p})$,
and $\varepsilon({\bf p})$ defined by Eq. (22).
As $g_2\to 0$, we have that
$\Delta({\bf p})\to 0$, and Eq. (25) becomes
\beq (\varepsilon({\bf p})-\mu) \tanh(2\alpha_{\bf p})=0.
\eeq Equation (26) requires that
\beq \varepsilon({\bf p})-\mu=0,\,\,\,\,
{\mathrm {if}}\,\,\, \tanh(2\alpha_{\bf p})\neq 0\,\,\,\, (0<n({\bf
p})<1), \eeq
which leads to the FC solutions defined by Eq. (4) \cite{vsl,vs}.
As soon as the coupling constant $g_2$ becomes finite but small, such
that $g_2/g_1\ll 1$, the plateau $\varepsilon({\bf p})-\mu=0$ is
slightly tilted and rounded off at the end points. This
implies that \beq
\varepsilon({\bf p})-\mu\sim \Delta_1,
\eeq
which allows us to estimate the effective mass as
\beq
\frac{M^*_{FC}}{M}\sim\frac{T_f}{\Delta_1}.
\eeq
Outside the condensate area, the quasiparticle dispersion
is determined by the effective mass $M^*_L$ given by Eq. (3). We
note that calculations in the context of a simple model
support the above consideration \cite{dkss}. In that case,
putting $V({\bf p}_1,{\bf p}_2)=\delta({\bf p}_1,{\bf p}_2)$
and $V_{pp}({\bf p}_1,{\bf p}_2)=\delta({\bf p}_1,{\bf p}_2)$
in Eqs. (19) and (20) and carrying out direct calculations,
we obtain at $T=0$
\beq E_0=\varepsilon(p_f)-\varepsilon(p_i)\simeq
\frac{(p_f-p_i)p_F}{M^*_{FC}}\simeq 2\Delta_1.
\eeq
On the other hand, at $T\geq T_c$, taking into account
that $n(p_i)\simeq1$ and $n(p_f)\simeq0$,
we obtain from Eq. (5) with the same accuracy,
\beq E_0\simeq\frac{(p_f-p_i)p_F}{M^*_{FC}}\simeq 2T.  \eeq
Equations (30) and (31) allow us to estimate the effective mass
$M^*_{FC}$ related to the region occupied by the FC at temperatures
$T\ll T_f$. Outside the region, the effective mass is
$M^*_L$. When Eqs. (28) and (29) are compared with Eqs. (5) and
(7), it is apparent that the gap $\Delta_1$ plays the role of the
effective temperature that defines the slope of the plateau. On the
other hand, at $T=T_c$ in OD or OP samples, the gap vanishes and
Eqs. (5) and (31) define the quasiparticle dispersion and the
effective mass. Taking into account that $\Delta_1\sim T_c$, we are
led to the conclusion that Eqs. (28) and (29) derived at $T=0$
match Eqs. (5) and (7) at $T_c$. Thus, Eqs. (28) and (29) are
approximately valid over the range $0\leq T\leq T_c$.  It follows
from Eq. (30) that at $T\leq T_c$, the quasiparticle dispersion can
be presented with two straight lines characterized by the respective
effective masses $M^*_{FC}$ and $M^*_L$  and
intersecting near the binding energy $E_0\sim 2\Delta_1$.
Equation (31) implies above $T_c$, the lines intersect near the
binding energy $\sim 2T$. The break separating the faster dispersing
high-energy part related to $M^*_L$ from the slower dispersing
low-energy part defined by $M^*_{FC}$ is likely to be enhanced in UD
samples at least because of the rise of the temperature $T_f$,
which grows with the decrease of doping.
We recall that in accordance with our assumption, the
condensate volume $\Omega_{FC}$ and $T_f$ are growing with
underdoping, see Eq. (6) and Sec. III. It was also suggested
that the FC arises near the van Hove singularities, while the
FC different areas overlap only slightly. Therefore, as one
moves from $(0,0)$ towards $(\pi,0)$ the ratio $M^*_{FC}/M^*_{L}$
grows in magnitude, developing into the distinct break.  In fact,
assuming that the temperature $T_f$ depends on the angle $\phi$ along
the Fermi surface and taking Eq. (29) into account,
one can arrive at the same conclusion. The dispersions above
$T_c$ exhibit the same structure except that the effective mass
$M^*_{FC}$ is governed by Eq. (31) rather than (30) and both the
dispersion and the break are partly ``covered" by the quasiparticle
width.  Thus, one concludes that there also exists a new energy
scale at $T\ll T_f$ defined by $E_0$ and intimately related to $T_f$
\cite{ars}.

We turn to the quasiparticle excitations with
the energy $E(\phi)=\sqrt{\varepsilon^2(\phi)+\Delta^2(\phi)}$.
At temperatures $T<T_c$, they are typical excitations of the
superconducting state. We now qualitatively
analyze the processes contributing to the width $\gamma$.  Within
the limits of the analysis, we can take $\Delta\simeq 0$,
which corresponds to considering excitations at the node. Our
treatment is then valid for both $T\leq T_c$ and $T_c\leq T$. For
definiteness, we consider the decay of a particle with the
momentum $p>p_F$. Then $\gamma(p,\omega)$ is given by the imaginary
part of the diagram shown in Fig. 3a, where the wiggly lines stand
for the effective interaction. Because of the unitarity,
diagram 3b which represents the real events can be used
to calculate the width \cite{rit}
\beq
\gamma(p,\omega)=2\pi\int
\left|\frac{V(q)}{\epsilon(q,-\omega_{pq})}\right|^2
n({\bf k})(1-n({\bf k+q}))\delta(\omega_{pq}+\omega_{kq})
\frac{d{\bf q}d{\bf k}}{(2\pi)^4},
\eeq
with $\epsilon(q,-\omega_{pq})$ being the
complex dielectric constant and $V(q)/\epsilon$
the effective interaction. Here, ${\bf q}$ and
$\omega_{kq}=\varepsilon({\bf k+q})-\varepsilon({\bf k})$
are the transferred momentum and energy, respectively, and
$\omega_{pq}=\omega-\varepsilon({\bf p-q})$
is the decrease in the quasiparticle energy
as the result of the rescattering processes:
the quasiparticle with the energy $\omega$ decays into a
quasihole $\varepsilon({\bf k})$  and two quasiparticles
$\varepsilon({\bf p-q})$  and $\varepsilon({\bf k+q})$.
The transferred momentum $q$ must satisfy the condition
\beq
p>|{\bf p-q}|>p_F.\eeq
Equation (32) gives the width as a function of $p$ and $\omega$;
the width of a quasiparticle with the
energy $\varepsilon(p)$ is given by
$\gamma(p,\omega=\varepsilon(p))$.

\begin{figure}[tbp]
\centerline{\epsfxsize=12cm
\epsfbox{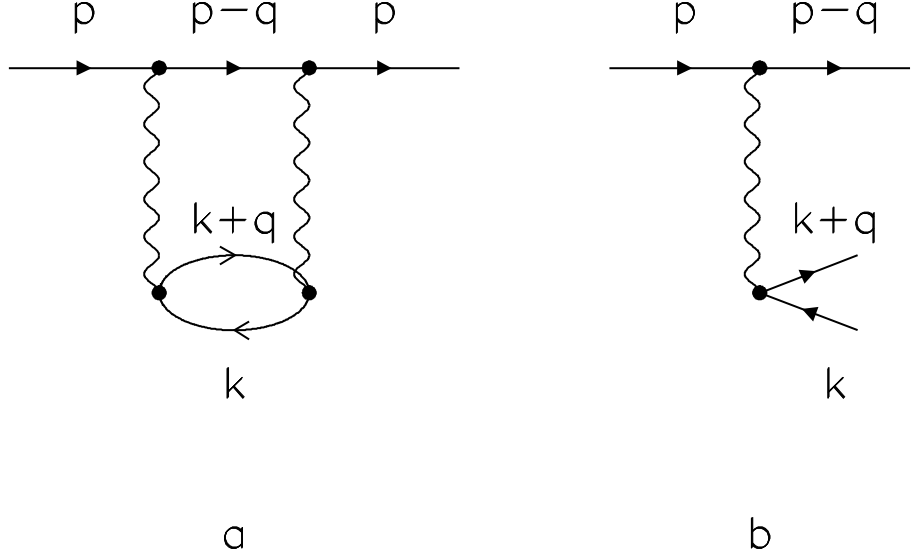}}
\caption{Diagram 3a depicts a process contributing to the imaginary
part. Diagram 3b shows a real process contributing to the imaginary
part, observe that quasiparticles ${\bf p-q}$, ${\bf k+q}$ and
${\bf k}$ are on the mass shell.}
\end{figure}

Estimating the width in Eq. (32)
with the constraint (33) and $\omega_{pq}\sim T$,
we find  that
\beq\gamma(p,\omega=\varepsilon(p))\sim (M^*_L)^3 T^2,\eeq
for normal Fermi liquids. In the case of the FC one could estimate
$\gamma\sim 1/T$ upon using Eqs. (9) and (34). This estimate were
correct if the dielectric constant is small, but
$\epsilon\sim M^*_{FC}$.  As the result, for the FC we have
\beq \gamma(p,\omega=\varepsilon(p))\sim
\frac{(M^*_{FC})^3 T^2}{(M^*_{FC})^2}\sim
T\frac{T_f}{\varepsilon_F},\eeq
where $\varepsilon_F$ is the
Fermi energy \cite{kss}. Calculating $\gamma(p,\omega)$ as a
function of $p$ at constant $\omega$, we obtain the same result for
the width given by Eq. (35) when $\omega=\varepsilon(p)$. The
calculated function can be fitted with a simple Lorentzian form,
because quasiparticles and quasiholes involved in the process
are also located in the vicinity of the Fermi level provided
$\omega-\varepsilon_F\sim T$.
It then follows from Eq. (35) that the
well-defined excitations exist at the Fermi surface even in the
normal state \cite{kss}. This result is in line with the
experimental findings determined from the scans at a constant
binding energy (momentum distribution curves or MDCs)
\cite{vall,vall1}. On the other hand, considering $\gamma(p,\omega)$
as a function of $\omega$ at constant $p$, we can check that the
quasiparticles and quasiholes contributing to the function can have
the energy of the same order of the magnitude. For
$\omega-\varepsilon_F\sim T$, the function is of the same Lorentzian
form, otherwise the shape of the function is
disturbed at high $\omega$ by high-energy excitations.
In that case the special form of the quasiparticle dispersion
characterized by the two effective masses must be taken into
account. As the result, the lineshape of the quasiparticle peak as a
function of the binding energy possesses a complex peak-dip-hump
structure \cite{blk,krc,kmf} directly defined by the existence of the
effective masses $M^*_{FC}$ and $M^*_L$.
Our consideration shows that it is the spectral peak obtained from
MDCs that provides important information on the existence of
well-defined excitations at the Fermi level and their width
\cite{ars}. The detailed numerical results will be presented
elsewhere.

At $T>T_c$, the gap is absent in OD or OP samples, and the width
$\gamma$ of excitations close to the Fermi surface is given by Eq.
(35). For UD samples, $\Delta(\theta)\equiv 0$ in the range
$\Omega_n$ and we have normal quasiparticle excitations with the
width $\gamma$. Outside the range $\Omega_n$, the Fermi level is
occupied by the BCS-type excitations with finite excitation
energy given by the gap $\Delta(\theta)$. Both types of excitations
have widths of the same order of magnitude. We now estimate $\gamma$.
For the entire Fermi level occupied by the normal state, the width is
equal to $\gamma\approx N(0)^3 T^2/\beta^2$, with the density of
states $N(0)\sim 1/T$ and the dielectric constant $\beta\sim N(0)$.
Thus, $\gamma\sim T$ \cite{dkss}. In our case, however only a part of
the Fermi level within $\Omega_n$ belongs to the normal excitations.
Therefore, the number of states allowed for quasiparticles and for
quasiholes is proportional to $\theta_c$, the factor $T^2$ is
therefore replaced by $T^2\theta_c^2$.  Taking these factors into
account, we obtain $\gamma\sim \theta_c^2 T\sim T(T-T_c)/T_c\sim
(T-T_c)$, because only small angles are considered.  Here, we have
omitted the small contribution coming from the BCS-type excitations.
That is why the width $\gamma$ vanishes at $T=T_c$.  Thus, the
foregoing analysis shows that in UD samples at $T>T_c$, the
superconducting gap smoothly transforms into the pseudogap. The
excitations of the gapped area of the Fermi surface have the same
width $\gamma\sim (T-T_c)$ and the region occupied by the pseudogap
is shrinking with increasing temperature.  These results are in good
qualitative agreement with the experimental facts $[4-7]$.

\section{Concluding remarks}
We have discussed the model of a strongly correlated
electron liquid based on the FC phase transition and extended it to
high-temperature superconductors. The FC transition
plays the role of a boundary separating the region of a strongly
interacting electron liquid from the region of a strongly correlated
electron liquid. On the basis of the BCS theory ideas we
have also considered the superconductivity with the d-wave symmetry
of the order parameter in the presence of the FC. We can conclude
that the BCS-type approach is fruitful for OD, OP, and UD
samples. We have shown that in
UD samples, the gap becomes flatter near the nodes at temperatures
$T<T_c$, and the superconducting gap smoothly
transforms into a pseudogap above $T_c$. The pseudogap occupies only
a part of the Fermi surface, which eventually shrinks with
increasing temperature, vanishing at $T=T^*$, and the maximum gap
$\Delta_1$ scales with the temperature $T^*$. We have also shown that
the general dependence of $T_c$, $T^*$, and $\Delta_1$ on the
underdoping level fits naturally into the considered model.  At
temperatures $T^*>T>T_c$, the single-particle excitations of the
gapped area of the Fermi surface have the width $\gamma\sim (T-T_c)$.
The quasiparticle dispersion in systems with FC can be
represented by two straight lines characterized by the
respective effective masses $M^*_{FC}$ and $M^*_L$.
At $T<T_c$, these lines intersect near the point $E_0\sim
2\Delta_1$, while above $T_c$, we have $E_0\sim 2T$. It is argued
that this strong change of the quasiparticle dispersion at $E_0$ can
be enhanced in UD samples because of strengthening the FC influence.
The single-particle excitations and their width $\gamma$ are also
studied.  We have shown that well-defined excitations with
$\gamma\sim T$ exist at the Fermi level even in the normal state.
This result is in line with the experimental findings determined from
the scans at a constant binding energy, or MDCs. We have also treated
the FC phase transition in the presence of the superconductivity and
shown that this phase transition can be considered as kinetic energy
driven. Thus, without any adjustable parameters, a number of the
fundamental problems of strongly correlated systems are naturally
explained within the proposed model.

This research was supported in part by the Russian Foundation for
Basic Research under Grant No. 98-02-16170.


\begin{thebibliography}{99}

\bibitem{sd} Z.-X. Shen and D.S. Dessau, Phys. Rep. {\bf 253}, 1
(1995).

\bibitem{ift} M. Imada, A. Fujimori, and Y. Tokura, Rev. Mod. Phys.
{\bf 70}, 1059 (1998).

\bibitem{d1} H. Ding {\it et al}., Nature {\bf 382}, 51 (1996).

\bibitem{d2} H. Ding {\it et al}., cond-mat/9712100.

\bibitem{camp2} M.R. Norman {\it et al}., cond-mat/9710163.

\bibitem{nr} M.R. Norman {\it et al}., cond-mat/9711232.

\bibitem{mn} J. Mesot {\it et al}., cond-mat/9812377 v2.

\bibitem{blk} P.V. Bogdanov {\it et al}.,
Phys. Rev. Lett. {\bf 85}, 2581 (2000).

\bibitem{krc} A. Kaminski {\it et al}., cond-mat/0004482.

\bibitem{vall} T. Valla {\it et al}., Science {\bf 285}, 2110 (1999).

\bibitem{ks} V.A. Khodel and V.R. Shaginyan,
JETP Lett. {\bf 51}, 553 (1990).

\bibitem{vol} G.E. Volovik, JETP Lett. {\bf 53}, 222 (1991).

\bibitem{lan} L.D. Landau, Sov. Phys. JETP {\bf 30}, 1058 (1956).

\bibitem{ksk} V.A. Khodel, V.R. Shaginyan, and V.V. Khodel,
Phys. Rep. {\bf 249}, 1 (1994).

\bibitem{dkss} J. Dukelsky {\it et al}.,
Z. Phys. {\bf 102}, 245 (1997).

\bibitem{vsl} V.R. Shaginyan, Phys. Lett. A {\bf 249}, 237 (1998).

\bibitem{kcs} V.A. Khodel, J.W. Clark, and V.R. Shaginyan,
Solid Stat. Comm. {\bf 96}, 353 (1995).

\bibitem{ksz} V.A. Khodel, V.R. Shaginyan, and M.V. Zverev,
JETP Lett. {\bf 65}, 253 (1997).

\bibitem{lp} M. Levy and J.P. Perdew, Phys. Rev. {\bf 48}, 11638
(1993).

\bibitem{sns} L. \'Swierkowski, D. Neilson, and J.
Szyma\'nski, Phys. Rev. Lett. {\bf 67}, 240 (1991).

\bibitem{scal} D.J. Scalapino {\it et al}., Phys. Rev.
B {\bf 34}, 8190 (1986).

\bibitem{scalr} D.J. Scalapino, Phys. Rep. {\bf 250}, 329 (1995).

\bibitem{abr} A.A. Abrikosov, Physica C {\bf 222}, 191 (1994);
A.A. Abrikosov, Phys. Rev. B {\bf 52}, R15738 (1995);
A.A. Abrikosov, cond-mat/9912394.

\bibitem{avl} M.Ya. Amusia and V.R. Shaginyan, Phys. Lett. A
{\bf 259}, 460 (1999).

\bibitem{vs} V.R. Shaginyan, JETP Lett. {\bf 68}, 527 (1998).

\bibitem{nm} N. Metoki {\it et al}., Phys. Rev. Lett.
{\bf 80}, 5417 (1998).

\bibitem{th} T. Honma {\it et al}., J. Phys. Soc. Jpn.
{\bf 68}, 338 (1999).

\bibitem{sp} J. Schmalian, D. Pines, and B. Stojkovic, Phys. Rev.
Lett.  {\bf 80}, 3839 (1998).

\bibitem{pr} I.A. Privorotsky, Sov. Phys. JETP {\bf 43}, 2255 (1962).

\bibitem{ov} A.W. Overhauser, Phys. Rev. {\bf 128}, 1437 (1962).

\bibitem{fong} H.F. Fong {\it et al}., Nature {\bf 398}, 588
(1999).

\bibitem{hhe} H. He {\it et al}., cond-mat/0002013 v2.

\bibitem{ab} A.A. Abrikosov, Phys. Rev. B {\bf 57}, 8656 (1998).

\bibitem{ncon} M.R. Norman {\it et al}., cond-mat/9912043;
M.R. Norman {\it et al}., cond-mat/0003406.

\bibitem{ch} G.V. Chester, Phys. Rev. {\bf 103}, 1693 (1956).

\bibitem{ars} S.A. Artamonov and V.R. Shaginyan,
JETP, in press.

\bibitem{rit} R.N. Ritchie, Phys. Rev. {\bf 114}, 644 (1959).

\bibitem{kss} V.A. Khodel, V.R. Shaginyan, and P. Schuck,
JETP Lett. {\bf 63}, 752 (1996).

\bibitem{vall1} T. Valla {\it et al}., Phys. Rev. Lett. {\bf 85}, 828
(2000).

\bibitem{kmf} A. Kaminski {\it et al}., Phys. Rev. Lett. {\bf 84},
1788 (2000).

\end{thebibliography}
\end{document}